\documentclass[fleqn,twoside]{article}
\usepackage{espcrc2}

\newcommand{\lsim}{\mathrel{\mathop{\kern 0pt \rlap
  {\raise.2ex\hbox{$<$}}}
  \lower.9ex\hbox{\kern-.190em $\sim$}}}
\newcommand{\gsim}{\mathrel{\mathop{\kern 0pt \rlap
  {\raise.2ex\hbox{$>$}}}
  \lower.9ex\hbox{\kern-.190em $\sim$}}}
\newcommand{\beq}     {\begin{equation}}
\newcommand{\eeq}     {\end{equation}}
\newcommand{\bea}     {\begin{eqnarray}}
\newcommand{\eea}     {\end{eqnarray}}
\newcommand{\intfive}      {\int d^5 x \sqrt{-G}}
\newcommand{\intfour}      {\int d^4 x }
\newcommand{\dt}      {\delta}

\newcommand{\n}      {{(n)}}

\newcommand{\es}      {\epsilon}
\newcommand{\ev}      {\equiv}

\newcommand{\lm}      {\lambda}

\newcommand{\sg}      {\sigma}

\newcommand{\Lg}       {{\mathcal L}}

\newcommand{\D}       {{\mathcal D}}
\newcommand{\no}      {\nonumber}

\usepackage{graphicx}
\usepackage[figuresright]{rotating}

\newcommand{\AmS}{{\protect\the\textfont2
  A\kern-.1667em\lower.5ex\hbox{M}\kern-.125emS}}

\title{Top quark Kaluza-Klein mode mixing in the Randall-Sundrum bulk \\
Standard Model and Constraint from $\Delta \rho$ }

\author{C. S. Kim\address{Department of Physics \& IPAP, Yonsei University, Seoul
120-749, Korea}%
\thanks{Supported
in part by Grant No. R02-2002-000-00168-0 from BRP of the KOSEF,
in part by Grant No. 2001-042-D00022 of the KRF.},
J. D. Kim\addressmark\thanks{Supported by Grant No. 2001-042-D00022 of the KRF.},
J.  Song\address{School of Physics,
Korea Institute for Advanced Study, Seoul 130-012, Korea}
}

\begin{document}

\begin{abstract}
The
Randall-Sundrum (RS) scenario with all the standard model (SM) fermions and gauge
bosons in the bulk is phenomenologically studied.
Even though the simple assumption of universal
bulk fermion mass $m_\psi$ leads to the same Kaluza-Klein (KK) mass spectrum for
all the SM fermions and thus suppresses new contributions to 
Flavor-Changing-Neutral-Current (FCNC)
and the $\rho$ parameter, large
Yukawa coupling of the top quark generates its KK mode mixing
and breaks the degeneracy: unacceptably large contribution
to $\Delta \rho$ occurs. With a different bulk fermion mass
to SU(2) singlet bottom quark, we demonstrate that there exists
some parameter space to satisfy the $\Delta\rho$ constraint.
\vspace{1pc}
\end{abstract}

\maketitle

\section{Introduction}

Extra dimensional models have drawn a lot of interest since it
can leave distinct
phenomenological signatures at future colliders.  In the original
RS scenario, the SM fields are assumed to be confined to our
brane. Phenomenological signatures come from KK gravitons with
electroweak scale masses and couplings to matter, characterized by
$\Lambda_\pi$. However, the small size of the RS-bulk allows that
the SM fields may also be in the bulk. In
Ref.\,\cite{Davoudiasl:1999tf}, it is demonstrated that placing
the SM gauge fields in the RS-bulk while confining the fermions to
our brane is strongly constrained by the current precision
electroweak data so that $\Lambda_\pi$ is pushed up to
about 100 TeV.
This is disfavored as a solution of the gauge
hierarchy problem. If both the SM gauge and fermion fields are in
the bulk \cite{Davoudiasl:2000wi}, their phenomenological
signatures are very sensitive to the bulk fermion mass $m_\psi$.

In the early study of the RS-bulk SM,
Yukawa interactions with the Higgs field have been ignored
due to small quark masses compared to the KK mass scale.
Then a simple assumption of universal bulk fermion mass
suppresses their contributions to
FCNC as well as
the $\rho$ parameter \cite{Davoudiasl:2000wi,Gherghetta:2000qt}.
This is because the common $m_\psi$ leads to the same KK mass spectra
for all the SM fermions.
The degeneracy of the up-type (and down-type) quark KK modes
operates the GIM cancellation \cite{GIM}
KK-level by level: With the
minimal flavor violation assumption that at the tree level the flavor mixing
comes only through the CKM matrix,
FCNC is suppressed as in the SM.
Since two constituents of SU(2)--doublet
have also the same KK mass,
their contribution to the $\rho$ parameter vanishes.
However, Yukawa interaction with the Higgs field
mixes the fermion KK tower members, which can be substantial for
the top quark  \cite{Davoudiasl:2000wi,top-mixing}.
Recently it has been shown that
the large mixing in the top quark KK sector leads to unacceptably large
contribution to the $\rho$ parameter and raises $\Lambda_\pi$
above 100 TeV  \cite{Hewett:2002fe}.
To accommodate those electroweak precision data,
a `mixed' scenario was proposed with
the third generation fermions on the TeV brane
but the other generations in the bulk.
However, the first excited KK mode of
gauge bosons should be heavier than 11 TeV
due to the strong constraints from precision measurements:
It is hard to probe the new physics effects at LHC.
In addition, the obvious discrimination of fermions according to generation
may lead to potentially dangerous FCNC due to the absence
of GIM mechanism.

It is worthwhile to keep the original framework
where all the fermions
are in the bulk, and to question other unsubstantiated  assumptions.
Relaxing
the universal bulk fermion mass assumption, we assign
a different bulk fermion mass $m_\psi'$ to
the SU(2)--singlet bottom quark field,
and see whether there exists some parameter space
to accommodate the $\rho$  constraint\cite{KimSong}.
We show that
this $m_\psi'$ allows some limited parameter space
where the degeneracy between the top and bottom quark KK modes is retained,
suppressing their contribution to the $\rho$ parameter.

\section{Extended Bulk SM in the RS scenario}
\label{review}

\subsection{Original set-up}
\label{GN}

Let us review the KK solution of a bulk fermion
with arbitrary Dirac bulk mass
in the RS scenario\cite{Gherghetta:2000qt,Grossman:1999ra,Chang:1999nh},
which causes a subtle problem
when discussing the bulk SM.
The five-dimensional action
of a Dirac fermion $\Psi$ with the bulk mass $m_\psi$
is
\bea
\label{action}
   S\!\!\!\!\!&=&\!\!\!\!\!\int\! \mbox{d}^4x \!\!\int\! \!\mbox{d}\phi \sqrt{-G}
   \bigg\{ E_{\underline{A}}^A \left[ \frac{i}{2} \bar\Psi\gamma^{\underline{A}}
   (D_A-\overleftarrow{D_A})\Psi
   \right] \nonumber \\
  & & - m_\psi \mbox{sign}(\phi) \bar\Psi\Psi \bigg\} . \nonumber
\eea
With the KK expansion of $\Psi$
\begin{equation}\label{KK}
   \Psi_{L,R}(x,\phi) = \sum_{n=0}^\infty
    \psi_{L,R}^\n(x)\,
   \frac{e^{2\sigma(\phi)}}{\sqrt{r_c}}\,\hat f_{L,R}^\n(\phi) \,,
\end{equation}
Eq.\,(\ref{action}) becomes the action
for a tower of massive Dirac fermions
\beq
\label{Sferm}
   S \!\!= \sum_{n=0}^\infty \int\!\mbox{d}^4x\,\Big\{
   \bar\psi^\n\,i\rlap/\partial\,\psi^\n
    - M_{f}^{(n)}\,\bar\psi^\n\,\psi^\n \Big\} \,.
\eeq
Note that $Z_2$-symmetric action constrains
$\bar\Psi\Psi=\bar\Psi_L\Psi_R+\bar\Psi_R\Psi_L$
 to be $Z_2$-odd.
If $\widehat{f}_L^{(n)}$ is a $Z_2$-even function $\chi^{(n)}$ then
$\widehat{f}^\n_{R}$ should be a $Z_2$-odd function $\tau^{(n)}$ and vice
versa. The order one parameter $\nu \ev m_\psi/k$ determines the
$\chi^{(n)}$ and $\tau^{(n)}$ as some combination of Bessel functions.

Let us discuss the physical implication of the parameter $\nu$.
Note that the canonically re-scaled zero mode of $Z_2$-even bulk fermion
is
proportional to $e^{ (1/2 +\nu)k r_c |\phi|}$.
For $\nu \!\! \ll \!\!-1/2$ the fermion bulk wave functions
are localized toward the Planck
brane:
The magnitudes of its gauge couplings with KK gauge bosons
are quite small.
Numerically for $\nu \lsim -0.5$ the couplings are too small
to be probed at high energy colliders\cite{Gherghetta:2000qt,Davoudiasl:2000my}.
For $\nu \!\!\gg \!\! 1$
the SM fermions become localized closer to the TeV brane:
The model approaches the RS model with the gauge fields only
in the bulk, which is phenomenologically disfavored\cite{Davoudiasl:1999tf}.
To be specific, if $\nu \gsim -0.3$,
the large contribution to the precision electroweak data
pushes $M_A^{(1)}$
up to about 6~TeV,
beyond the direct production at any planned collider.
Therefore, we consider the parameter space of $\nu$
between $-0.5$ and $-0.3$.

Another subtle point when placing
the SM fermions in the AdS$_5$ bulk is that
the fermion field contents should be doubled.
In the SM, a fermion field with left-handed chirality and
that with right-handed chirality belong to
different representations of a gauge group.
In the RS-bulk SM, however,
if a fermion which belongs to a specific representation of a gauge group
has a single chirality,
the bulk wave function cannot be determined.
For each generation,
we introduce four five-dimensional Dirac fields, an SU(2)--doublet
fermion field $Q=(q_u, q_d)^T$
and two SU(2)--singlet fermion fields,
$u$ and $d$,
with weak hypercharges
$Y=1/6$, $2/3$, and $-1/3$ respectively.

Since the SM fermion should correspond to the KK zero mode,
we assign $Z_2$-even wave function $\chi^\n$
to the left-handed SU(2)--doublet
and the right-handed SU(2)--singlet,
$u(x,\phi)$ and $d(x,\phi)$.
The charged current interactions,
mediated by the bulk $W$ boson,
connect $q_u$ and $q_d$:
\bea
S\!\!\!\!
& \supset &  \!\!\!\!
\intfour
\frac{g}{\sqrt{2}}
\sum_{l=0}^\infty
\left[ \sum _{n,m=0}^\infty
\bar{q}_{uL}^\n \rlap/W^{+(l)} q_{dL}^{(m)}
\left\{
       C_{nml}^{\bar{f}f' W}
\right\}
\right. \no \\
& &\hspace{-1cm} +\left.\sum _{n,m=1}^\infty
\bar{q}_{uR}^\n \rlap/W^{+(l)} q_{dR}^{(m)}
\left\{
       H_{nml}^{\bar{f}f' W}
\right\}\right]+h.c., \no
\eea
where $g ={g_5}/{\sqrt{2\pi r_c}} $.
$C_{nml}^{\bar{f}f' W}$ and
$H_{nml}^{\bar{f}f' W}$ denote the couplings of the $m$-th and
the $n$-th fermion states to the $l$-th $W$ boson
in the unit of the SM coupling. It is defined by
\bea
\label{C}
C_{nml}^{\bar{f}f' W}\!\!\!\!
&=&\!\!\!\!\sqrt{2\pi}\int_{-\pi}^\pi d\phi \, e^\sg \chi^\n(\phi)
\chi^{(m)}(\phi)\chi_A^{(l)}(\phi), \nonumber  \\
H_{nml}^{\bar{f}f' W}\!\!\!\!
&=&\!\!\!\! \sqrt{2\pi}\int_{-\pi}^\pi d\phi \,e^\sg \tau^\n
\tau^{(m)}\chi_A^{(l)} . \no
\eea

\subsection{Minimally Extended Model}
\label{our-model}

Now every SM fermion possesses its KK tower.
We remind the reader that in the RS background the fermion KK mass spectrum
is determined by the bulk fermion mass $m_\psi$.
A simple assumption of universal bulk fermion mass
results in the same KK mass spectrum for all the SM fermions.
However, there is another mass source,
Yukawa interaction.
This Yukawa mass relates SU(2)--doublet with singlet
(e.g., $m_{Y}\overline{q}_{uL}u_{R}$).
It is to be compared to the KK mass which is the coupling of
$\overline{q}_{uL} q_{uR}$ and $\overline{u}_{L} u_{R}$.
This difference results in mixing among the fermion KK modes.
Since quark masses are much smaller than the KK mass scale,
this mixing effect has been neglected  in the early study.
One exception is the top quark.
Its heavy mass yields substantial mixing.
The mass shift of the top quark KK mode from the rest up-type quark KK modes
invalidates the GIM cancellation;
FCNC becomes inevitable.
More severe problem happens if precision measurements
are taken into account.
In particular, the $\rho$ parameter becomes dangerous
since it does not follow the decoupling theorem since its quantum correction
increases with the squared mass difference between the $T=1/2$ and
$T=-1/2$ fermions.
As each top quark KK mass deviates from the
corresponding $b$ quark KK mass,
their additive contribution yields disastrous and unacceptable
value of $\Delta \rho$.
The problem does not ameliorate but worsens as we add
more and more KK states\cite{Hewett:2002fe}.

In Ref.\,\cite{Hewett:2002fe},
a `mixed' scenario was proposed such that
the third generation fermions are confined on the TeV brane
while the other two generations propagate in the bulk.
Instead, here we keep the original set-up but relax
the unsubstantiated assumption of the universal bulk fermion mass\cite{KimSong}.
For the simplest extension, we assume that
the SU(2)--singlet bottom quark field has different bulk fermion mass $m_\psi'$,
and see whether this introduction of another parameter
can accommodate the $\Delta\rho$ constraint without a new hierarchy.
Then the five dimensional action for the
third generation quarks becomes
\bea \label{actionSM}
   S \!\!\!\!\vspace{-1.1cm} &=&\!\!\!\!\vspace{-1.1cm}
\int\!\!\mbox{d}^4x\!\!\int\!\mbox{d}\phi\,\sqrt{-G}
   [ E_a^A i(
   \bar{Q}\gamma^a \D_A Q
 + \bar{t}\gamma^a \D_A t   \no \\
 &+&\vspace{-8.0cm}\!\!\!\!\bar{b}\gamma^a \D_A b)
    - \mbox{sign}(\phi)\left(
    m_\psi\{ \bar{Q}Q + \bar{t}t\}
    +m_\psi'\,\bar{b}b \right)] \,. \no
\eea
The
KK mass terms become
\bea
\label{KK-mass}
\Lg &=&- \sum_{n=1}^\infty k_{EW} [
x_f^\n(\nu)\{ \bar{q}^\n_{t L} q_{tR}^\n +\bar{q}^\n_{b L}
q_{b R}^\n  \no  \\
 &+& \bar{t}^\n_L t_R^\n \}
 +{x_f^\n}(\nu\,') \bar{b}^\n_L b_R^\n ] + h.c.,
\eea
where we introduced additional dimensionless parameter $\nu\,'\!\!=\!m_\psi'/k$.
As explicitly shown in Eq.\,(\ref{KK-mass}),
the KK masses of fermions depend on the
bulk fermion masses, $m_\psi$ and $ m_\psi'$.

The five-dimensional action for Yukawa interaction with the
confined Higgs field is
\bea \label{KK-f-mass}
S_{ffH}\!\!\!&=&\!\!\!
-\intfive [  \frac{\lm_5^b}{k} \,{\overline{Q}}(x,\phi) \cdot H(x) b
(x,\phi)  \no \\
 & & \hspace{-1.5cm} +\frac{\lm_5^t}{k} \,\es^{ab}
\overline{Q}(x,\phi)_a \cdot  H(x)_b t(x,\phi)
+h.c. ] \dt(\phi-\pi) \,, \no
\eea
where $\lm_5^{b,t}$ is the
five-dimensional Yukawa coupling.
Spontaneous symmetry breaking shifts the Higgs
field as $H^0 \to v_5 +H^{'0}$ with a VEV of the order Planck
scale.

\section{Constraints from the $\rho$ parameter}
\label{rho-constraint}

The $\rho$ parameter
is defined by the difference between the $W$ and $Z$ boson self-energy
functions re-scaled by each mass.
It
has been known to
play a special role among
precision measurements
since it is sensitive to heavy fermions beyond SM.
With the Higgs mass below 1 TeV,
the current electroweak precision data constrain
$\Delta\rho<2\times 10^{-3}$  at 95\% CL
(with $\Delta\rho \equiv \rho-\rho_{\rm SM}$),
which restricts new SU(2)--doublet
fermion mass spectrum to satisfy  \cite{PDG}
\begin{equation}
\label{rho}
\sum_i \Delta m_i^2 \le (115 \,\mbox{GeV})^2 .
\end{equation}
Note that for $|M_t^\n-M_b^\n| \ll M_t^\n$,
we have
\beq
\Delta m^2 \approx \sum_n (M_t^\n-M_b^\n)^2
\,.
\eeq

Thus different mass spectra for the top and bottom quark KK modes
lead to dangerous contributions to $\Delta\rho$.
First let us demonstrate that the RS-bulk SM with the universal bulk fermion
mass assumption ($\nu=\nu'$)
cannot satisfy this $\Delta\rho$ constraint.
\begin{figure}
\includegraphics[height=7.5cm,angle=-90]{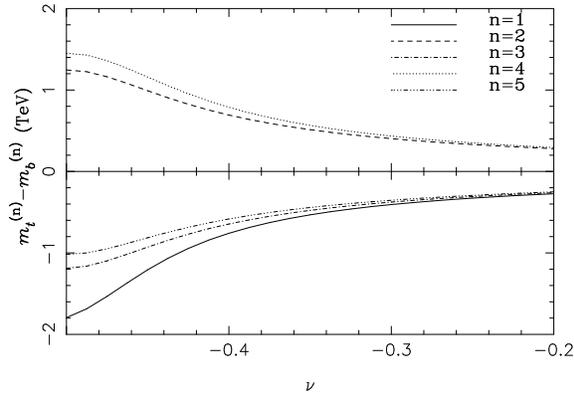}
\caption{\label{deltamuni} The KK mass difference
between the top and bottom quarks
when $\nu=\nu'$ and $k_{EW}=1$ TeV.
}
\end{figure}
In Fig.\,\ref{deltamuni}, we show
top and bottom KK mass differences up to the fifth KK excitation
states as a function of parameter $\nu$,
which is too large to satisfy Eq.\,(\ref{rho}).

\begin{figure}
\includegraphics[height=7.5cm,angle=-90]{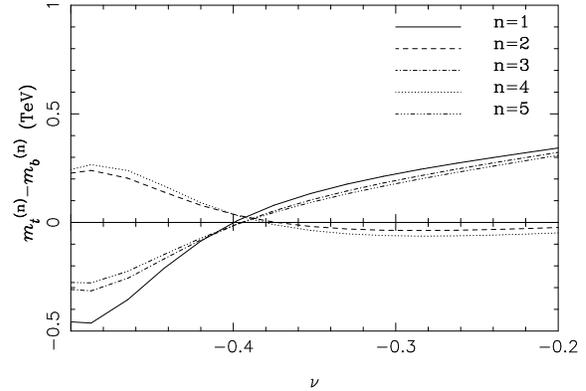}
\caption{\label{deltam} The KK mass difference
between the top and bottom quarks
when $\nu'=-0.6$ and $k_{EW}=1$ TeV.
}
\end{figure}

Now  let us allow $\nu \neq\nu'$
and see whether there exists a parameter space
to satisfy Eq.\,(\ref{rho}).
In Fig.\,\ref{deltam}, we
shows, with the fixed $\nu'=-0.6$, the
top-bottom quark mass differences for the first five excited
modes. We find that for example the $\nu=-0.39$ case with $\nu\,'=-0.6$
gives vanishing
mass differences below 20 GeV for the first five KK excited
states.
A remarkable point is that
the $\Delta m$ decreases for higher KK
modes; the contribution of higher KK modes becomes less important.

In summary,
the minimal RS-bulk SM with a common bulk fermion mass
has disastrous contribution to the $\rho$ parameter
due to
quite large mass shifts between the top and bottom quark KK modes.
We relax the universal bulk fermion mass,
and let the SU(2)--singlet bottom quark field have
a different bulk fermion mass $m_\psi'$.
It is shown that for example if $m_\psi'/k \simeq -0.6$
and $m_\psi/k \simeq -0.4$,
the degeneracy of the top and bottom quark KK mode is good enough
to suppress the new contribution to $\Delta \rho$.

\def\MPL #1 #2 #3 {Mod. Phys. Lett. {\bf#1},\ #2 (#3)}
\def\NPB #1 #2 #3 {Nucl. Phys. {\bf#1},\ #2 (#3)}
\def\PLB #1 #2 #3 {Phys. Lett. {\bf#1},\ #2 (#3)}
\def\PR #1 #2 #3 {Phys. Rep. {\bf#1},\ #2 (#3)}
\def\PRD #1 #2 #3 {Phys. Rev. {\bf#1},\ #2 (#3)}
\def\PRL #1 #2 #3 {Phys. Rev. Lett. {\bf#1},\ #2 (#3)}
\def\RMP #1 #2 #3 {Rev. Mod. Phys. {\bf#1},\ #2 (#3)}
\def\NIM #1 #2 #3 {Nucl. Inst. Meth. {\bf#1},\ #2 (#3)}
\def\ZPC #1 #2 #3 {Z. Phys. {\bf#1},\ #2 (#3)}
\def\EJPC #1 #2 #3 {E. Phys. J. {\bf#1},\ #2 (#3)}
\def\IJMP #1 #2 #3 {Int. J. Mod. Phys. {\bf#1},\ #2 (#3)}
\def\JHEP #1 #2 #3 {J. High En. Phys. {\bf#1},\ #2 (#3)}

\end{document}